\documentstyle[aps,prb,multicol,epsf]{revtex}

\begin{document}

\title{Cu-O network dependent core-hole screening 
       in low-dimensional cuprates: A theoretical analysis}

\author{C.~Waidacher, J.~Richter, and K.~W.~Becker}

\address{Institut f\"ur Theoretische Physik,
  Technische Universit\"at Dresden, D-01062 Dresden, Germany}
  
\maketitle

\begin{abstract}
We analyze the influence of the dimensionality on the Cu $2p_{3/2}$
core level X-ray photoemission spectra of undoped cuprates. Both exchange 
splitting and delocalization properties are described within one framework 
using a multi-band Hubbard model. The spectral intensity is calculated 
by means of the Mori-Zwanzig projection technique for a 
CuO$_{4}$ plaquette ('zero'-dimensional), an infinite CuO$_{3}$ chain
(one-dimensional), and an infinite CuO$_{2}$ plane (two-dimensional). The
results are compared to the spectra of Bi$_{2}$CuO$_{4}$, Sr$_{2}$CuO$_{3}$,
and Sr$_{2}$CuO$_{2}$Cl$_{2}$. Our analysis allows to distinguish between 
effects of the Cu-O geometry and effects from material-specific properties. 
\end{abstract}

\begin{multicols}{2}

%%%%%%%%%%%%%%%%%%%%%%%%%%%% SECTION I %%%%%%%%%%%%%%%%%%%%%%%%%%%%%%%%%%

\section{Introduction}

The influence of dimensionality and lattice geometry on the
electronic structure of crystals is a classical topic in solid state
physics. However, in the presence of strong electronic correlations like
those observed in the high-$T_{c}$ superconducting cuprates, the theoretical
analysis of electronic properties is especially difficult. Recently, the
electronic structure of various undoped cuprates has been studied
experimentally using core-level X-ray photoemission spectroscopy 
(XPS).~\cite{Boeske97,Boeske98} By comparing these spectra to the results 
of theoretical model calculations we can try to improve our understanding 
of the electronic properties of the cuprates. In addition, a theoretical
analysis allows us to check the validity of the microscopic models used for
the calculations, and it provides values for the physical parameters which
are contained in the model Hamiltonians.

The core-level photoemission process leads to a final state in which an
electron is missing in the $2p_{3/2}$ core orbital of a Cu site. The
resulting positive charge is screened by valence electrons, predominantly
from Cu $3d_{x^{2}-y^{2}}$ and O $2p_{x,y}$ orbitals. The contribution of
these screening processes to the final state determines the detailed form of
the experimental spectrum. Therefore, core-level spectra contain information
about the dynamics of valence electrons.

Since the atomic ground state of
the Cu-O system is a $2p^{6}3d^{9}$ configuration, it is advantageous to use
the hole picture. In this case the core-level photoemission process amounts
to the creation of a core hole which has a strong repulsive Coulomb
interaction with the valence hole in the $3d_{x^{2}-y^{2}}$ orbital of the
same Cu site. 

Experimentally, the Cu $2p$ core-level spectrum of
many formally divalent Cu compounds shows a dominant main line 
and a pronounced satellite 
structure.\cite{vanderLaan81,Fujimori87,Shen87,Ghijsen88}
In analogy to an approach for metals, \cite{Kotani73} this satellite structure
was assigned  to a (``poorly screened'') final state (denoted $3d^{9}$) in 
which the valence hole largely remains on the Cu site.\cite{Larsson77} 
In this way, the structure of the 
satellite could be explained by multiplet splitting due to the remaining 
$d$-hole.\cite{Okada89,Goldoni94,vanVeenendaal94,Tanaka97}

The main line was interpreted as originating from a final state 
in which the hole resides in the neighbouring ligands (denoted 
$3d^{10}$\underline{L}). However, for several compounds the
main line is found to be asymmetric, and to have a large width.
For instance, three substructures were identified in the main line 
of Bi$_{2}$Sr$_{2}$CaCu$_{2}$O$_{8}$\cite{Parmigiani91} and 
Sr$_{2}$CuO$_{2}$Cl$_{2}$ \cite{Boeske97,Boeske98} 
(which will be analyzed in this paper). 
Substructures of this kind are the reason why the main line in the 
Cu $2p_{3/2}$ spectra cannot be attributed to a local $3d^{10}$\underline{L} 
excitation only: This interpretation does not account for the asymmetry and 
the large width of the main line. 

An alternative explanation was found using numerical exact diagonalization
for a chain of three plaquettes 
(Cu$_{3}$O$_{10}$).\cite{vanVeenendaal93prl,vanVeenendaal93}
It was shown that the valence hole could delocalize further in the 
crystal. This leads to a lowest eigenstate of $3d^{10}$ character, in which 
the hole is mainly pushed out onto the neighbouring CuO$_{4}$ units, forming a
Zhang-Rice singlet.\cite{Zhang88} These
delocalization processes in the hole picture are equivalent to the screening
processes in the electron picture.

Obviously, this screening of the core hole depends on the geometry of the 
CuO network.\cite{Okada97jesrp} This suggests the possibility of 
finite-size effects in cluster calculations.\cite{Okada95}
In fact, convergence of the spectra with respect to the system size was 
found only for chain clusters with lengths of seven plaquettes 
(Cu$_{7}$O$_{21}$).\cite{Okada96}
These calculations led to a satisfactory agreement with the 
experimental results for Sr$_{2}$CuO$_{3}$.\cite{Okada96}

Another popular model for the description of core-level spectroscopy is
the single-site Anderson impurity model.\cite{Gunnarsson83a,Gunnarsson83b}
Recently, the influence of dimensionality on Cu $2p_{3/2}$ spectra of
several cuprates was analyzed using this model.\cite{Karlsson99}
Overall, very good agreement with the experimental
main lines was obtained. The satellite structures, however, were not
discussed. In particular, no multiplet effects were included. Furthermore,
the absolute energetical positions of the main lines of Sr$_{2}$CuO$_{3}$ and
Sr$_{2}$CuO$_{2}$Cl$_{2}$ were not correctly reproduced. These shortcomings
are avoided in our treatment.

In the present paper, we discuss the influence of 
of the dimensionality on the Cu $2p_{3/2}$ core level spectra of undoped 
cuprates by means of the Mori-Zwanzig projection technique. Both exchange 
splitting and delocalization properties are described within one framework 
using a multi-band Hubbard model. In addition, a clear description of the 
final states is obtained.

Our analysis allows to distinguish between 
effects of the Cu-O geometry and effects from material-specific properties.
Examples for material-specific properties are the Cu-O distance
or components which do not belong to the Cu-O network. Some of these 
properties are described by the values of model parameters like the Cu-O 
charge-transfer energy or the Cu-O hybridization strength.
A phenomenological analysis cannot conclusively determine whether an effect is
due to dimensionality or not. For instance, due to differences in
material-specific properties a peak which is brought about by a similar
process in two different compounds may not appear at the same binding energy
in both spectra. Therefore, the investigation of microscopic models is
necessary.

We compare our calculations with the experimental spectra 
of three materials which consist of different Cu-O networks: 
Bi$_{2}$CuO$_{4}$ which contains separated CuO$_{4}$ plaquettes~\cite{Ong90} 
(``zero''-dimensional Cu-O$_{4}$ network), Sr$_{2}$CuO$_{3}$ where the 
plaquettes form linear chains~\cite{Ami95} (one-dimensional Cu-O$_{3}$ 
network), and Sr$_{2}$CuO$_{2}$Cl$_{2}$ which contains CuO$_{2}$ 
planes~\cite{Miller90} (two-dimensional Cu-O$_{2}$ network). 

In the case of zero-dimensional edge-shared or separated 
plaquettes (like Bi$_{2}$CuO$_{4}$) the main line consists of a single 
feature which is explained by a local process in which the valence hole 
moves from the core-hole site to the surrounding O 
sites.~\cite{Okada96,Waidacher99b} 
For one-dimensional zigzag or linear chains (like Sr$_{2}
$CuO$_{3}$) two features contribute to the main line. In the case
of two-dimensional planar systems (like Sr$_{2}$CuO$_{2}$Cl$_{2}$) the main
line is composed of at least three features. In addition to these effects
there is an interesting trend in the intensity ratio $I_{s}/I_{m}$ between
the satellite and the main line. For zero-dimensional systems this ratio is
rather large, e.g.\ $I_{s}/I_{m}\leq 0.58$ for Bi$_{2}$CuO$_{4}$. On the
other hand, in the case of one-dimensional networks the ratio is small, for
instance $I_{s}/I_{m}=0.37$ for Sr$_{2}$CuO$_{3}$. The ratio for
two-dimensional systems lies in-between these values, e.g. $I_{s}/I_{m}=0.52$
for Sr$_{2}$CuO$_{2}$Cl$_{2}$.

The paper is organized as follows. In Sec.\ II we introduce the
model Hamiltonian and describe the method of calculation. Section III
contains a discussion of the results. Our conclusions are summarized in Sec.\ 
IV.

%%%%%%%%%%%%%%%%%%%%%%%%%%%% SECTION II %%%%%%%%%%%%%%%%%%%%%%%%%%%%%%%%%%

\section{Model and calculation}

For the calculation of the Cu 2p$_{3/2}$ photoemission spectra we use a
multi-band Hubbard Hamiltonian to describe the Cu-O network, while
additional terms represent the interaction between the core hole and valence
holes on the core-hole site. The full Hamiltonian in the hole picture is
\begin{eqnarray}
H &=&\Delta\sum_{j\sigma}n^p_{j\sigma}
+U_{dd} \sum_{i}n^d_{i\uparrow}n^d_{i\downarrow}
+U_{dc} \sum_{\sigma\xi}n^d_{0\sigma}n^c_{0\xi}\nonumber\\
&&\quad+I_{dc}~\mbox{\bf S}^d_{0}\cdot\mbox{\bf J}^c_{0}
+t_{pd} \sum_{\langle ij\rangle\sigma}\phi^{ij}_{pd}
(p^\dagger_{j\sigma}d_{i\sigma}+h.c.)\nonumber\\
&&\quad+t_{pp} \sum_{\langle jj^\prime\rangle\sigma}\phi^{jj^\prime}_{pp}
p^\dagger_{j\sigma}p_{j^\prime\sigma}~\mbox{,}\label{Hamilton}
\end{eqnarray}
where $d^\dagger_{i\sigma}$ ($p^\dagger_{j\sigma}$) create a hole 
with spin $\sigma$ in the $i$-th Cu $3d$ orbital ($j$-th O $2p$ 
orbital) and $n^d_{i\sigma}$ ($n^p_{j\sigma}$) are the 
corresponding occupation-number operators. The first and second term 
on the r.h.s.\ of Eq.~(\ref{Hamilton}) describe 
the charge-transfer energy $\Delta$ and 
the on-site Coulomb repulsion $U_{dd}$ between Cu $3d$ 
valence holes. The third and fourth term represent the local Coulomb 
repulsion $U_{dc}$ and the effective exchange interaction $I_{dc}$ 
between the $3d$ valence holes and the $2p_{3/2}$ core hole at the 
core-hole site (which is taken to be site $i=0$). 
$\mbox{\bf S}^d_{0}$ is the spin-$1/2$ operator of a Cu $3d$ hole, 
while $\mbox{\bf J}^c_{0}$ and $n^c_{0\xi}$ are the pseudo-spin 
$3/2$ operator and the number operator of the $2p_{3/2}$ core hole 
(with $\xi =\pm3/2,~\pm1/2$) at site $i=0$. Finally, the last 
two terms on the r.h.s.\ of Eq.~(\ref{Hamilton}) describe 
the hybridization of Cu $3d$ and O $2p$ orbitals (hopping strength 
$t_{pd}$) and of O $2p$ orbitals (hopping strength $t_{pp}$). The 
factors $\phi^{ij}_{pd}$ and $\phi^{jj^\prime}_{pp}$ give the correct 
sign for the hopping processes and $\langle ij\rangle$ denotes the
summation over nearest neighbor pairs. Hamiltonian (\ref{Hamilton}) describes
delocalization processes and multiplet splitting within one framework. 

The spectral intensity $I\left( \omega \right)$, as a function of binding
energy $\omega$, is obtained from the hole-hole correlation function 
\begin{eqnarray}
I\left( \omega \right) &=&-\sum_{\xi }\Im \left[ G_{00}^{\xi }\left(
\omega +i0\right) \right]  \label{Intensity}~\mbox{,} \\
G_{00}^{\xi }\left( \omega +i0\right) &=& \langle\Psi|c_{0\xi}~ 
\frac{1}{\omega +i0- {\cal L}} ~c^\dagger_{0\xi}|\Psi\rangle
~\mbox{,}  \label{Hole-hole-correlation}
\end{eqnarray}
where $c^\dagger_{0\xi}$ creates a core hole with pseudo-spin $\xi$ 
at site $i=0$. ${\cal L}$ is the Liouville operator defined by 
${\cal L}A= [H,A]$ for any operator $A$. $|\Psi\rangle$ is the full ground 
state of $H$ before the core hole is created in the photoemission process.

As illustrated in Ref.~\onlinecite{Waidacher99a}, the full ground state 
$\left| \Psi \right\rangle$ can be constructed by starting from the atomic, 
N\'{e}el-ordered ground state $\left| \psi _{0}\right\rangle$. 
In $\left| \psi _{0}\right\rangle$ 
all O sites are empty and every Cu site is singly occupied with 
alternating spin direction. Using an exponential transformation, 
$\left| \psi_{0}\right\rangle$ is approximately transformed into the 
full ground state $\left| \Psi \right\rangle$
\begin{equation}
\label{Ground-state}
|\Psi\rangle = \exp \left(\sum_{i\alpha}\lambda_{\alpha}F_{i,\alpha}\right)
|\Psi_N\rangle~\mbox{,}
\end{equation}
with fluctuation operators $F_{i,\alpha }$ and fluctuation strengths 
$\lambda _{\alpha }$. The operators $F_{i,\alpha }$ describe delocalizations
of a valence hole which was initially located at Cu site $i$. For instance,
fluctuation $F_{i,2d}$ describes the creation of doubly occupied Cu sites
due to the hopping of a hole from Cu site $i$, via nearest neighbor O site 
$j$, to the nearest neighbor Cu sites $k$
\[
F_{i,2d} =\sum_{jk\sigma }n_{k\overline{\sigma }}^{d}~
d_{k\sigma }^{\dagger }\left( 1-n_{j\sigma }^{p}\right) 
d_{i\sigma }~\mbox{.}
\]
In the case of a single CuO$_{4}$ plaquette one fluctuation operator is
sufficient to obtain the exact ground state. For a CuO$_{3}$ chain we use up
to nine fluctuation operators thus allowing for delocalizations leading as
far as to the next-nearest neighbor plaquette. Due to its higher symmetry,
the ground state of a CuO$_{2}$ plane is well described by only five 
fluctuation operators. For a detailed description of the fluctuation operators 
$F_{i,\alpha }$ see Ref.~\onlinecite{Waidacher99a}. The fluctuation 
strengths $\lambda _{\alpha }$
are determined using the set of equations 
\begin{equation}
0=\left\langle \Psi \right| \left[ H,F_{\alpha }^{\dagger }\right] \left|
\Psi \right\rangle \text{ , }\alpha =1,2,\ldots\quad\mbox{.}  
\label{lambda-equations}
\end{equation}
which follows from the condition that $\left| \Psi \right\rangle$ is an
eigenstate of the full Hamiltonian (\ref{Hamilton}). The parameters 
$\lambda_{\alpha }$ are found to decrease exponentially with increasing 
length of the fluctuation processes. This result is a retrospective 
justification for the neglect of fluctuations which lead beyond the 
range covered by the operators $F_{i,\alpha}$. Ground state 
(\ref{Ground-state}) has charge properties which are size consistent and 
agree well with the results of Quantum Monte Carlo 
simulations.~\cite{Waidacher99a}

From Eq.(\ref{Intensity}) the Cu $2p_{3/2}$ photoemission spectra are
calculated using the Mori-Zwanzig projection technique.~\cite{Mori65} 
This method uses a set of operators $D_{\mu}$, the so-called dynamical 
variables. For these dynamical variables the following matrix equation 
holds 
\begin{equation}
\sum_{\gamma }\left( z\delta _{\mu \gamma }-\omega _{\mu \gamma }-\Sigma
_{\mu \gamma }\left( z\right) \right) G_{\gamma \nu }\left( z\right) =\chi
_{\mu \nu }~\mbox{,}  \label{Projection-equation}
\end{equation}
where $z=\omega +i0$, and where $\delta _{\mu \gamma }$ is the unity matrix.
The correlation functions $G_{\gamma \nu}\left( z\right)$ are given by 
\begin{equation}
G_{\gamma \nu }\left( z\right) =\left\langle \Psi \right| D_{\gamma
}^{\dagger }\frac{1}{z-\mathcal{L}}D_{\nu }\left| \Psi \right\rangle
~\mbox{.}\label{Correlation-function}
\end{equation}
In Eq.(\ref{Projection-equation}) the susceptibility matrix 
$\chi _{\mu \nu}$, the frequency matrix $\omega _{\mu \gamma }$, 
and the self-energy matrix $\Sigma _{\mu \gamma }$ are defined by 
\begin{eqnarray}
\chi _{\mu \nu } &=&\left\langle \Psi \right| D_{\mu }^{\dagger }D_{\nu
}\left| \Psi \right\rangle~\mbox{,}  \label{Susceptibility} \\
\omega _{\mu \gamma} &=&\sum_{\eta }\left\langle \Psi \right| 
D_{\mu}^{\dagger }{\mathcal L}D_{\eta}\left| \Psi \right\rangle 
\chi _{\eta \gamma}^{-1}~\mbox{,}  \label{Frequency} \\
\Sigma _{\mu \gamma }\left( z\right) &=&\sum_{\eta}\left\langle \Psi
\right| D_{\mu }^{\dagger }{\mathcal L}Q \frac{1}{z-Q{\mathcal L}Q}Q
{\mathcal L}D_{\eta }\left| \Psi \right\rangle \chi _{\eta \gamma }^{-1}
~\mbox{.}\label{Self-energy}
\end{eqnarray}
$Q$ is a projector on the subspace which is orthogonal to the space spanned
by the dynamical variables $D_{\mu}$. It is defined by 
\begin{equation}
Q=1-\sum_{\mu \nu }D_{\mu }\left| \Psi \right\rangle \chi _{\mu \nu
}^{-1}\left\langle \Psi \right| D_{\nu }^{\dagger }~\mbox{.}  
\label{Q-projector}
\end{equation}
The set $\left\{ D_{\mu }\right\} $ should contain the dynamical variable 
$D_{0\xi }=c_{0\xi }^{\dagger }$. Then, one of the correlation functions in 
Eq.(\ref{Correlation-function}) is the hole-hole correlation function 
$G_{00}^{\xi }$ of Eq.(\ref{Hole-hole-correlation}). In this case, 
solving Eq.(\ref{Projection-equation}) for $G_{00}^{\xi }$ one obtains 
the spectral intensity $I\left( \omega \right)$ from Eq.(\ref{Intensity}). 
One possible approach to obtain an approximate solution of 
Eq.(\ref{Projection-equation}) is to make the set 
$\left\{ D_{\mu }\right\}$ of dynamical variables sufficiently large so 
that the self-energies $\Sigma _{\mu \gamma }$ can be neglected. 
In this case an approximation for 
$I\left( \omega \right)$ results which can be systematically improved by 
further enlarging the set of dynamical variables until the results are 
converged. Due to the neglect of the self-energy matrix the calculated 
spectra show no broadening. Thus, a convolution with an artificial line 
width $\Gamma$ is necessary for a comparison with the experiment.

\epsfxsize=0.4\textwidth
\epsfbox{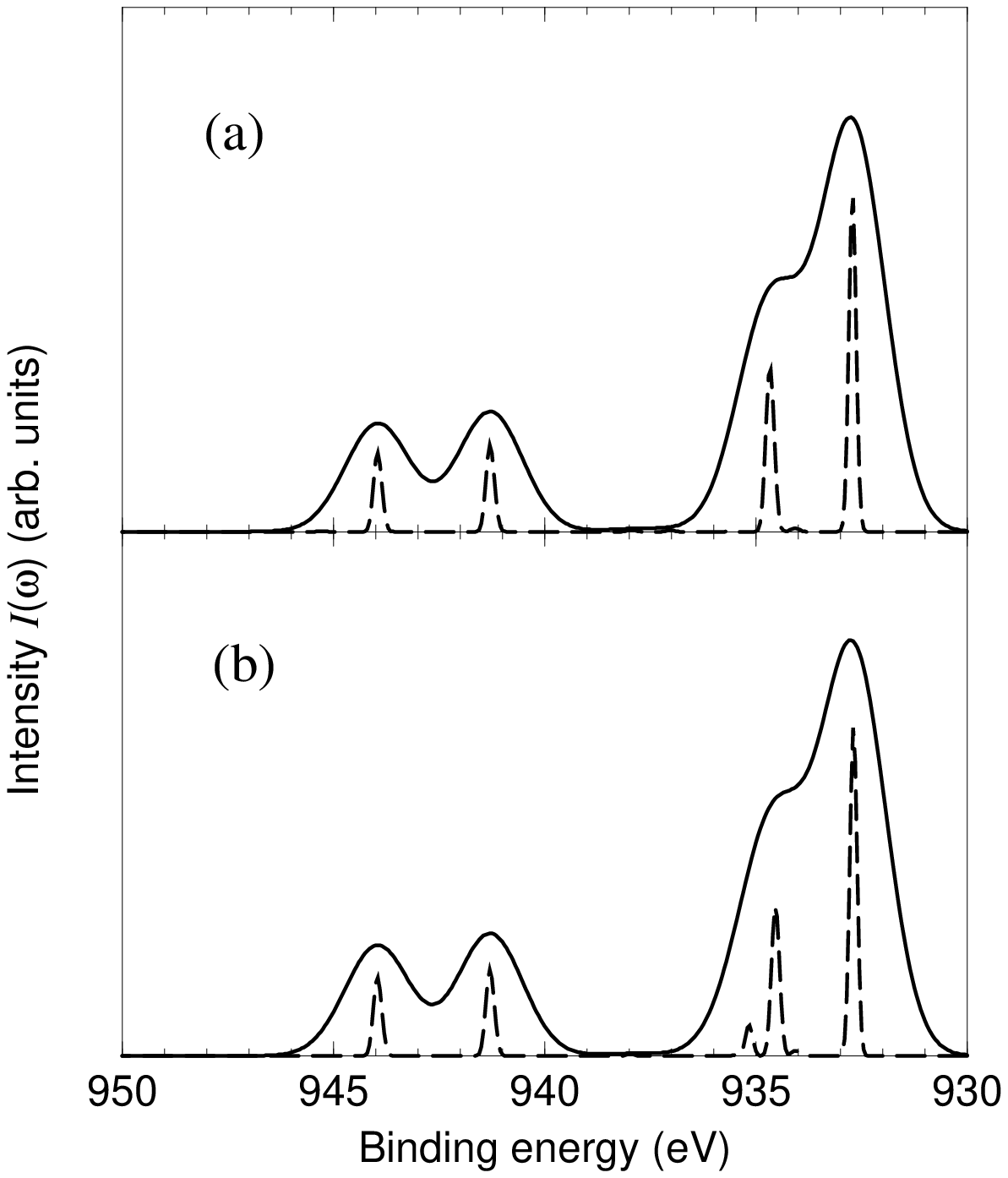}
\begin{figure}
\narrowtext
\caption{Convergence of the projection technique. The calculated spectra 
for an infinite CuO$_{3}$ chain are shown convoluted with a Gaussian 
function of width 
$\Gamma =1.8~\mbox{eV}$ (solid lines) and $\Gamma =0.2~\mbox{eV}$ (dashed 
lines). (a) and (b) show the result with $40$ and $70$ dynamical variables, 
respectively. The additional dynamical variables used in the second case 
describe far-reaching delocalization processes leading to the next-nearest 
neighbor plaquette. Their inclusion leads to a small redistribution of 
spectral weight around $935~\mbox{eV}$ binding energy which is only visible
for the smaller line width $0.2~\mbox{eV}$.}
\end{figure}

Besides $D_{0\xi}$ the chosen set of dynamical variables includes the operator 
\[
D_{0\xi }^{^{\prime }}=\sum_{\sigma }d_{0,-\sigma }^{\dagger }d_{0\sigma
}c_{0,\xi +2\sigma }^{\dagger }~\mbox{,}
\]
which describes the creation of a Cu core hole on site $i=0$ and a spin flip 
of the valence hole on the same site due to the exchange interaction with the
core hole. Furthermore, we include the following dynamical variables 
\begin{eqnarray}
D_{\alpha \xi } &=&F_{0,\alpha }c_{0\xi }^{\dagger }~\mbox{,} \nonumber \\
D_{\alpha \xi }^{^{\prime }} &=&F_{0,\alpha }\sum_{\sigma }d_{0,-\sigma
}^{\dagger }d_{0\sigma }c_{0,\xi +2\sigma }^{\dagger }~\mbox{,} \nonumber
\end{eqnarray}
for all fluctuation operators $F_{0,\alpha }$ used in the ground-state
calculation. These dynamical variables describe the creation of a core hole
and the subsequent delocalization of the valence hole from the same site,
without $\left( D_{\alpha \xi }\right) $ and with spin flip
$\left( D_{\alpha\xi }^{^{\prime }}\right)$. In the case of a single 
CuO$_{4}$ plaquette a set of $14$ dynamical variables suffices for the 
exact solution. For a CuO$_{2}$ plane $33$ variables lead to a well-converged 
spectrum, while $40$ variables are sufficient for a CuO$_{3}$ chain. 

In Fig.1 the convergence of the results is exemplified for the case of a 
CuO$_{3}$ chain. Figure 1(a) shows the spectrum obtained with $40$ 
dynamical variables while for the spectrum in Fig. 1(b) the set of 
dynamical variables is almost twice as large ($70$ variables). The 
inclusion of the additional variables leads to a small redistribution of 
spectral weight around $935$ eV binding energy. This change is only visible 
when a small line width $\Gamma$ is used for the convolution of the line 
spectrum.

%%%%%%%%%%%%%%%%%%%%%%%%%%%% SECTION III %%%%%%%%%%%%%%%%%%%%%%%%%%%%%%%%%%

\section{Results}

A typical set of values for the parameters in model (\ref{Hamilton}) 
has been obtained for La$_{2}$CuO$_{4}$ by band-structure 
calculation~\cite{McMahan88}
\begin{eqnarray}
\Delta&=&3.5\text{ eV, }U_{d}=8.8\text{ eV,} \nonumber \\
t_{pd}&=&1.3\text{ eV, }t_{pp}=0.65\text{ eV.} \label{Values}
\end{eqnarray}
For the comparison with the experimental spectra we start with this 
set and adapt some of the parameters to the specific systems which
we investigate.
The value of the exchange parameter $I_{dc}=-1.5$ eV and the line width 
$\Gamma =1.8~\mbox{eV}$ of the Gaussian function to convolute the spectrum 
are obtained from a comparison of the exact solution for a single CuO$_{4}$
plaquette with the experimental result for Bi$_{2}$CuO$_{4}$, see Fig.~2. 
From this comparison we furthermore determine the absolute energetical
position of all calculated spectra. Thereby we allow for general shifts
of $\pm0.3~\mbox{eV}$ which is the experimental accuracy of the
absolute energy values.~\cite{Boeske98}

To fit the experimental spectra, we vary $\Delta$ and $U_{dc}$ until the 
calculated satellite to main-peak intensity ratio $I_{s}/I_{m}$ coincides 
with the experimental one. It is not possible to use only $\Delta$ as
fit parameter because the separation between the satellite and the main 
line is mainly determined by the difference $U_{dc}-\Delta$. However, we can
reduce the number of free parameters by keeping this difference constant.

\epsfxsize=0.4\textwidth
\epsfbox{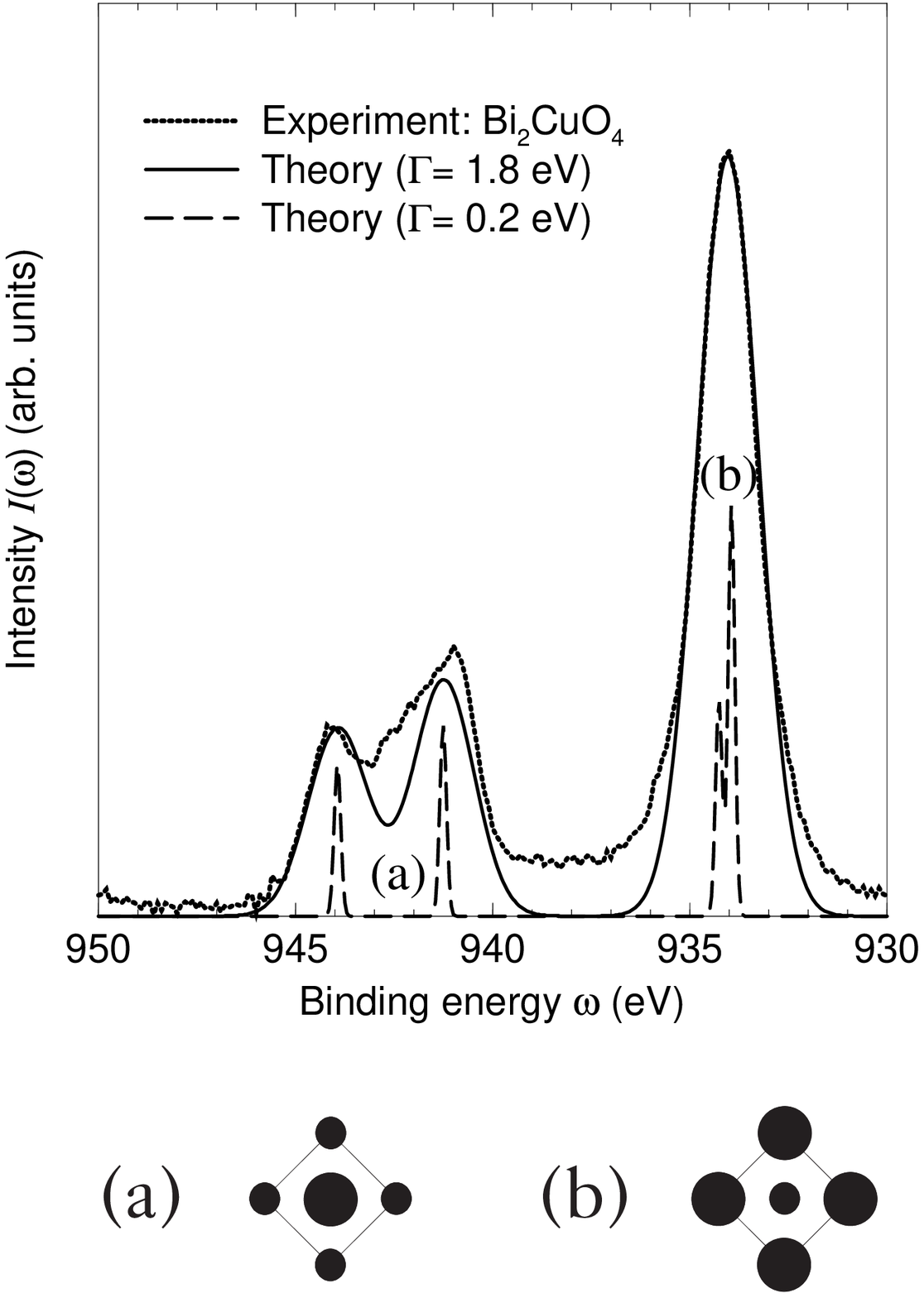}
\begin{figure}
\narrowtext
\caption{Bi$_{2}$CuO$_{4}$: Comparison of experimental data (dots) from
Ref.~\protect\onlinecite{Boeske98} with the result of the projection technique 
for a single plaquette with $\Delta =3.5~\mbox{eV}$. The theoretical 
spectra have been convoluted with a Gaussian function of width 
$\Gamma =1.8~\mbox{eV}$ (solid line) and $\Gamma =0.2~\mbox{eV}$ 
(dashed line). The valence hole delocalization in the final states is 
shown below. The core-hole site is the central Cu site. Large (small) 
dots symbolize a large (small) density of the valence hole originally 
located at the core-hole site. The final states (a) and (b) are 
associated to the corresponding lines in the spectrum.}
\end{figure}

For the spectra of Bi$_{2}$CuO$_{4}$ and Sr$_{2}$CuO$_{3}$ we keep 
$U_{dc}-\Delta =5~\mbox{eV}$, while the smaller satellite-main line 
separation in the spectrum of Sr$_{2}$CuO$_{2}$Cl$_{2}$ is taken into 
account by keeping $U_{dc}-\Delta =4.2~\mbox{eV}$. In this way, we obtain 
a good fit of the Bi$_{2}$CuO$_{4}$ spectrum with $\Delta =3.5~\mbox{eV}$, 
see Fig.~2, while the spectrum of Sr$_{2}$CuO$_{3}$ is well described by 
the result of the projection technique for a CuO$_{3}$ chain with 
$\Delta =2.7~\mbox{eV}$, see Fig.~3. These results have already been 
described elsewhere.~\cite{Waidacher99b} Here, we therefore discuss only 
those aspects which 
are relevant for the influence of dimensionality and lattice geometry. 

The lower parts of Figs.~2 and 3 show the delocalization properties of 
the final states which lead to the most important lines in the calculated 
spectra. The final state associated with the satellite lines, i.e.\ state 
(a) in Figs.~2 and 3, is highly localized. Since the valence hole remains 
mainly on the core-hole site, the influence of geometry is negligible. 
This is consistent with the experimental observation that the satellite 
peaks depend much less on the geometry than the main lines. 

\epsfxsize=0.4\textwidth
\epsfbox{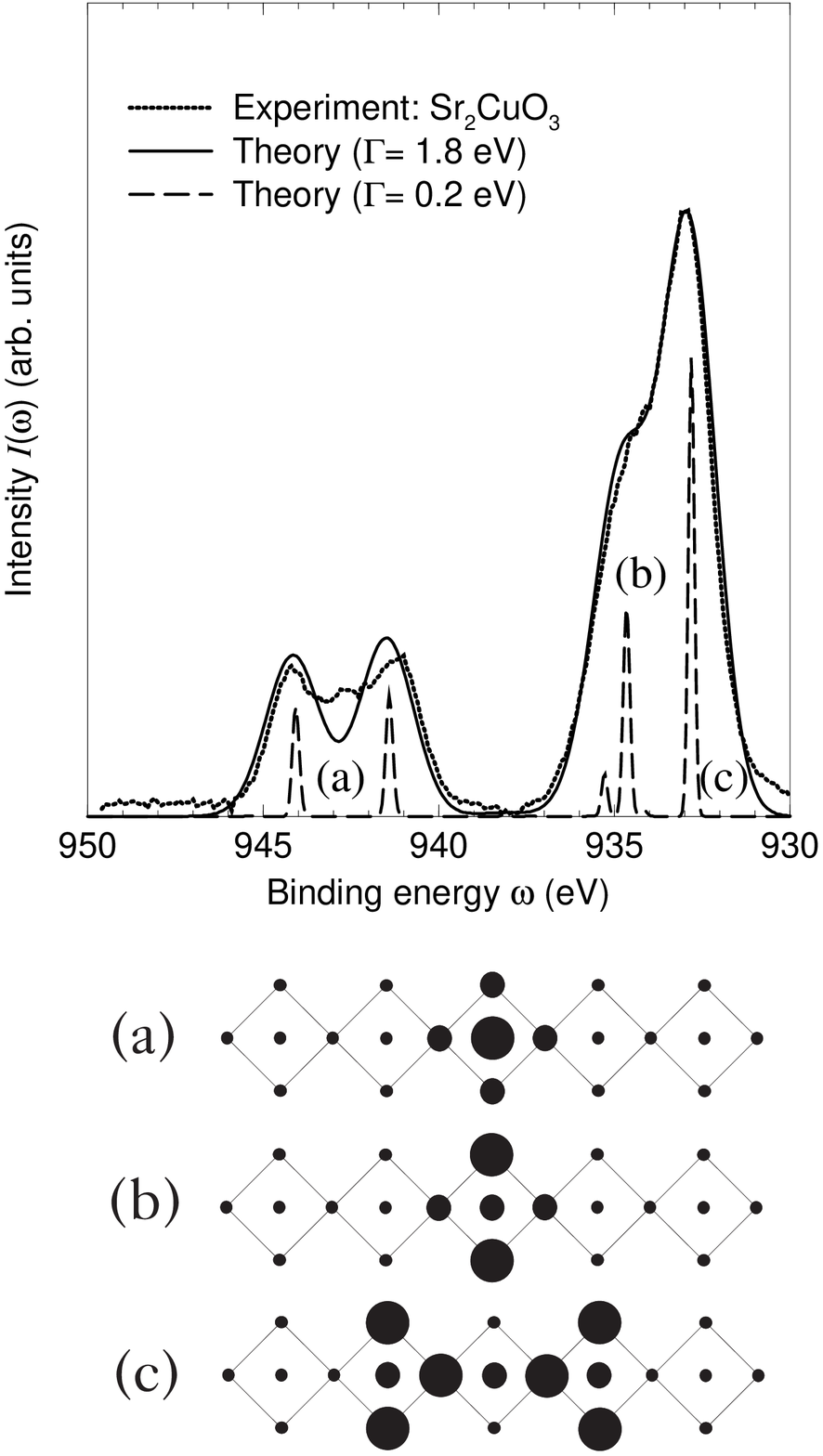}
\begin{figure}
\narrowtext
\caption{Sr$_{2}$CuO$_{3}$: Comparison of experimental data (dots) from
Ref.~\protect\onlinecite{Boeske98} with the result of the projection 
technique for an infinite 
CuO$_{3}$ chain with $\Delta =2.7~\mbox{eV}$. The line width is 
$\Gamma =1.8~\mbox{eV}$ (solid line) and $\Gamma =0.2~\mbox{eV}$ (dashed 
line). The valence hole delocalization in the final states is shown below. 
The core-hole site is the Cu site in the central plaquette. Large (small) 
dots symbolize a large (small) density of the valence hole 
originally located at the core-hole site. The final states (a), (b) and (c) 
are associated to the corresponding lines in the spectrum.}
\end{figure}

Final state (b) which causes the main line in the CuO$_{4}$ plaquette spectrum 
(Fig.~2) leads to a shoulder in the main line of the CuO$_{3}$ 
chain spectrum (Fig.~3) and is also rather localized, with most of the 
valence hole density concentrated on the O sites surrounding the core-hole 
site. However, in the case of the CuO$_{3}$ chain the hole density at 
the O sites in chain direction is smaller than at the O sites 
perpendicular to it. A comparable effect has been observed in exact 
diagonalization studies.~\cite{Okada96,Okada95,Okada98} 
The most important effect of the 
network geometry is the emergence of peak (c) in the CuO$_{3}$ chain 
spectrum, see Fig.~3. This peak, which dominates the main line, is 
associated with a delocalization of the valence hole to the neighboring 
plaquettes which may be interpreted as the formation of a Zhang-Rice
singlet.~\cite{Okada95,vanVeenendaal93}

\epsfxsize=0.4\textwidth
\epsfbox{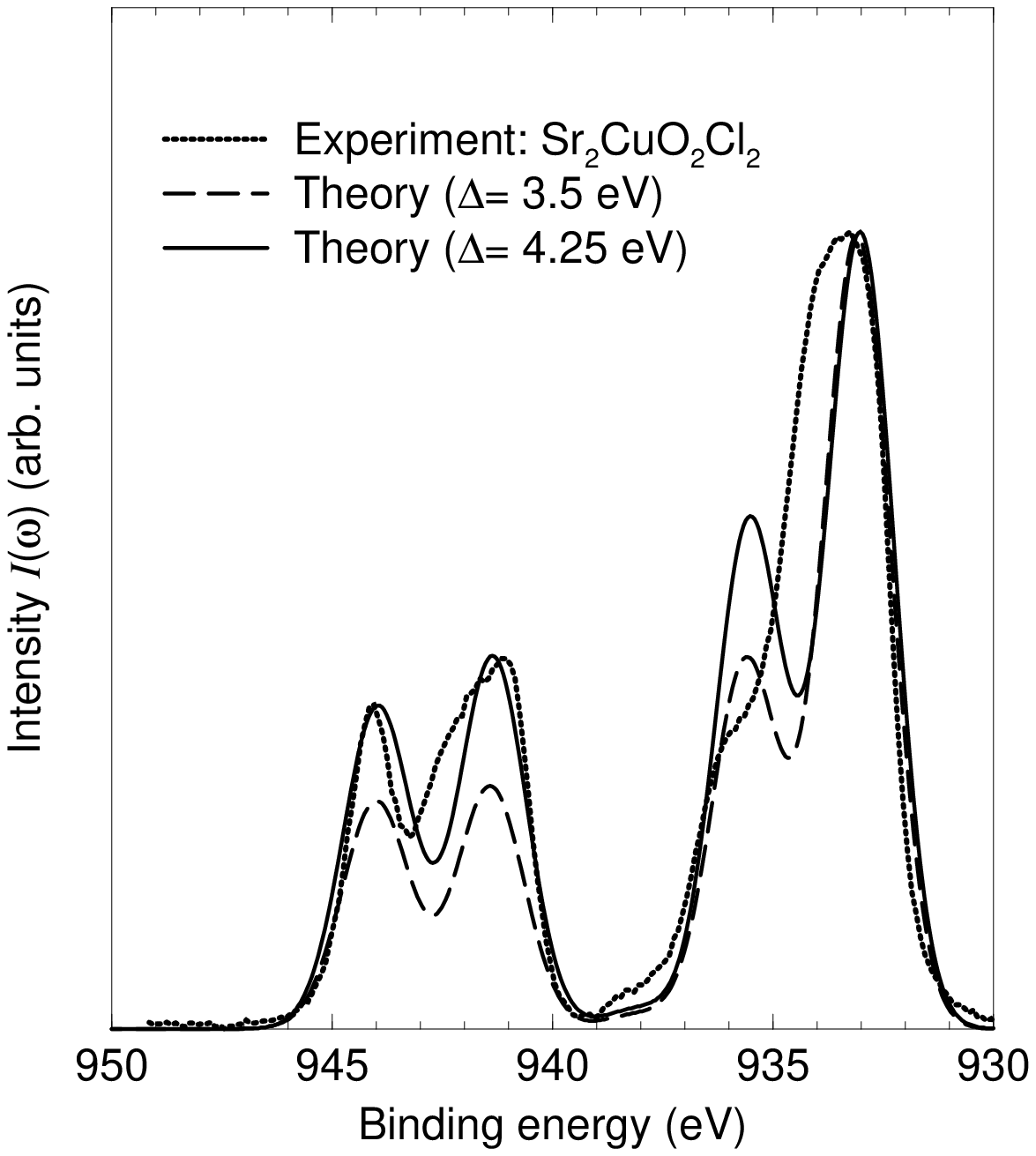}
\begin{figure}
\narrowtext
\caption{Sr$_{2}$CuO$_{2}$Cl$_{2}$: Comparison of experimental data (dots) 
from Ref.~\protect\onlinecite{Boeske98} with the result of the projection 
technique 
for an infinite CuO$_{2}$ plane. $\Delta$ is used as a fit parameter: 
$\Delta =3.5~\mbox{eV}$ (dashed line) and $\Delta =4.25~\mbox{eV}$ 
(solid line), all other parameter values are those of set (\ref{Values}).
The line width is in both cases $\Gamma =1.8~\mbox{eV}$. 
For $\Delta =4.25~\mbox{eV}$ the 
intensity ratio $I_{s}/I_{m}$ between satellite and main line is equal 
to the experimental one, but the form of the main line is not reproduced 
correctly.}
\end{figure}

In the case of the CuO$_{2}$ plane it turns out that a fit 
using $\Delta$ as described above does not lead to a 
satisfying agreement with the spectrum of 
Sr$_{2}$CuO$_{2} $Cl$_{2}$, see Fig.~4. For the standard value 
$\Delta =3.5~\mbox{eV}$ (dashed line in Fig.~4) the calculated ratio 
$I_{s}/I_{m}=0.4$ is too small compared to the experimental value 
$I_{s}/I_{m}=0.52$. With increasing $\Delta$ (i.e.\ with decreasing charge 
fluctuations) the relative intensity of both the satellite and the 
shoulder structure around $935~\mbox{eV}$ binding energy increases. For 
$\Delta =4.25~\mbox{eV}$ (solid line in Fig.~4) the calculated ratio 
$I_{s}/I_{m}$ is equal to the experimental one. However, the form of 
the main line is not reproduced correctly. The intensity of the 
shoulder structure is overestimated while a line seems to be missing 
around $934~\mbox{eV}$ binding energy.

In order to obtain a better fit we have varied $t_{pd}$ and $U_{dc}$
instead of $\Delta$ and $U_{dc}$. As shown in Fig.~5 this leads to a 
better agreement with the spectrum of Sr$_{2}$CuO$_{2}$Cl$_{2}$.
Keeping $\Delta =3.5~\mbox{eV}$, the experimental ratio 
$I_{s}/I_{m}=0.52$ is reproduced for $t_{pd}=1.15~\mbox{eV}$
and $U_{dc}=8.1~\mbox{eV}$. The delocalization properties of the 
most important final states are shown in the lower part of Fig.5. 
As opposed to the zero- and one-dimensional cases shown in Figs.~2 
and 3, the valence hole from the core-hole site may now delocalize 
in two dimensions. Nevertheless the final states obtained for a 
CuO$_{2}$ plane are rather similar to those of the lower dimensional 
systems. For state (a) this is easy to explain by the local 
nature of the satellite peak. The fact that the delocalization in 
states (b) and (c) is not significantly larger in two dimensions than in 
one dimension is, on the other hand, somewhat surprising. Key element in 
the explanation of this effect are the four Cu sites which are the 
diagonal nearest neighbors of the core-hole site (cf.\ the lower part 
of Fig.~5). Due to antiferromagnetic correlations these diagonal 
Cu sites are predominantly
occupied by valence holes which have the same spin direction as the 
valence hole on the core-hole site. Therefore, due to the Pauli principle 
the holes from the diagonal Cu sites suppress fluctuations of the valence 
hole from the core-hole site. The increase of charge fluctuations due to 
the higher dimensionality is largely compensated by this suppression. An 
analogous effect has already been found to be important for the 
ground-state charge properties of Hamiltonian 
(\ref{Hamilton}).~\cite{Waidacher99a}

It is interesting to compare the results of the projection technique 
with exact diagonalization calculations (without multiplet splitting) 
of a Cu$_{5}$O$_{16}$ cluster from 
Refs.~\protect\onlinecite{Boeske98,Okada97jesrp}. This 
cluster contains five plaquettes in a cross-like configuration where 
the central Cu site is the core-hole site. Notice that the 
Cu$_{5}$O$_{16}$ system does not contain the diagonal Cu sites which,
as discussed above, suppress fluctuations from the central Cu site. 
Therefore one expects this system to display features of artificially 
strong delocalization like, e.g., a reduced ratio $I_{s}/I_{m}$.
In fact, while the diagonalization shows a similar shoulder structure 
as the projection technique, cf.\ Fig.~4, the intensity of this 
shoulder, its separation from the lowest-energy line as well as the 
ratio $I_{s}/I_{m}$ are smaller.

The slightly reduced value $t_{pd}=1.15~\mbox{eV}$ for the Cu-O hopping 
strength which leads to the spectrum shown in Fig.~5 may be explained by 
the larger Cu-O distance in Sr$_{2}$CuO$_{2}$Cl$_{2}$ compared to 
La$_{2}$CuO$_{4}$.~\cite{Miller90} Nevertheless, the agreement of the 
theoretical result with the experimental spectrum is still not satisfactory. 
The calculated main line is dominated by two features, and at least 
one excitation seems to be missing in the region around $934~\mbox{eV}$ 
binding energy. Notice, that in the theoretical main line of
Sr$_{2}$CuO$_{2}$Cl$_{2}$ from Ref.~\protect\onlinecite{Karlsson99} an 
excitation is missing as well.

These results suggest the conclusion that Hamiltonian 
(\ref{Hamilton}) still does not include all degrees of freedom which are 
necessary for a detailed description of the main line in the Cu 2p$_{3/2}$ 
spectrum of Sr$_{2}$CuO$_{2}$Cl$_{2}$. Since all effects of  the Cu-O 
network dimensionality are already included in model (\ref{Hamilton}) we 
expect the missing third main-line feature to be a material-specific 
effect. Orbitals which are not yet taken into account 
are, for example, non-planar orbitals in the CuO$_{2}$
system, like the Cu $3d_{z^{2}-r^{2}}$ orbital. However, in view of the 
good agreement with the experiments shown in Figs.~2 and 3, these orbitals 
do not seem necessary for the description of Bi$_{2}$CuO$_{4}$ and 
Sr$_{2}$CuO$_{3}$. 

\epsfxsize=0.4\textwidth
\epsfbox{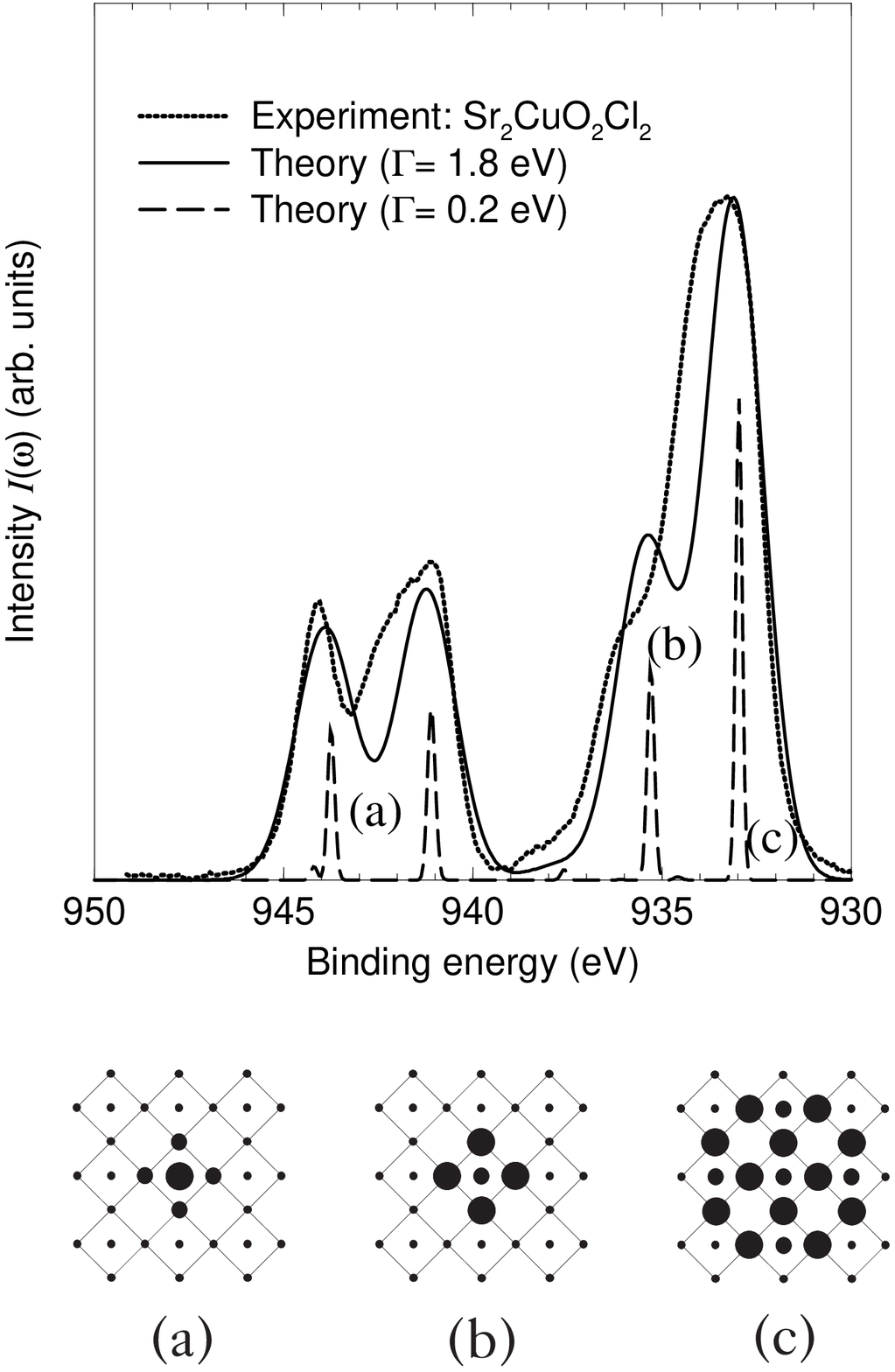}
\begin{figure}
\narrowtext
\caption{Sr$_{2}$CuO$_{2}$Cl$_{2}$: Comparison of experimental data 
(dots) from Ref.~\protect\onlinecite{Boeske98} with the result of the 
projection technique for an infinite CuO$_{2}$ plane. $t_{pd}$ is used
as a fit parameter. $t_{pd}=1.15~\mbox{eV}$ and $U_{dc}=8.1~\mbox{eV}$. 
The line width is 
$\Gamma =1.8~\mbox{eV}$ (solid line) and $\Gamma =0.2~\mbox{eV}$ (dashed 
line). The valence hole delocalization in the final states is shown below. 
The core-hole site is the Cu site in the central plaquette. Large (medium, 
small) dots symbolize a large (medium, small) density of the valence hole 
originally located at the core-hole site. The final states (a), (b) and (c) 
are associated to the corresponding lines in the spectrum.}
\end{figure}

It may also be possible that sites which do not belong 
to the CuO$_{2}$ plane (like the Cl apex site) contribute to the screening 
in Sr$_{2}$CuO$_{2}$Cl$_{2}$.
Notice that the Cu $2p$ main line in copper oxides without apex-oxygen 
site (CuO, Nd$_{2}$CuO$_{4}$) is narrower than in compounds with one apex 
oxygen (Bi$_{2}$Sr$_{2}$CaCu$_{2}$O$_{8}$) or two apex oxygens 
(La$_{2}$CuO$_{4}$) per copper site.\cite{Parmigiani92}
 However, both the large spatial distance 
between the Cl and the CuO$_{2}$ plane and the large energy difference 
between the Cl-$3p$ and the Cu-$3d$-O-$2p$ line in the valence 
photoelectron spectrum~\cite{Boeske97} suggest that screening from Cl 
sites should be small. Another possible candidate for the explanation 
of the missing third main-line feature are non-bonding oxygen $2p$ orbitals.
Recently, a feature in the optical spectrum of Sr$_{2}$CuO$_{2}$Cl$_{2}$
has been attributed to these orbitals.\cite{Choi99}

\epsfxsize=0.4\textwidth
\epsfbox{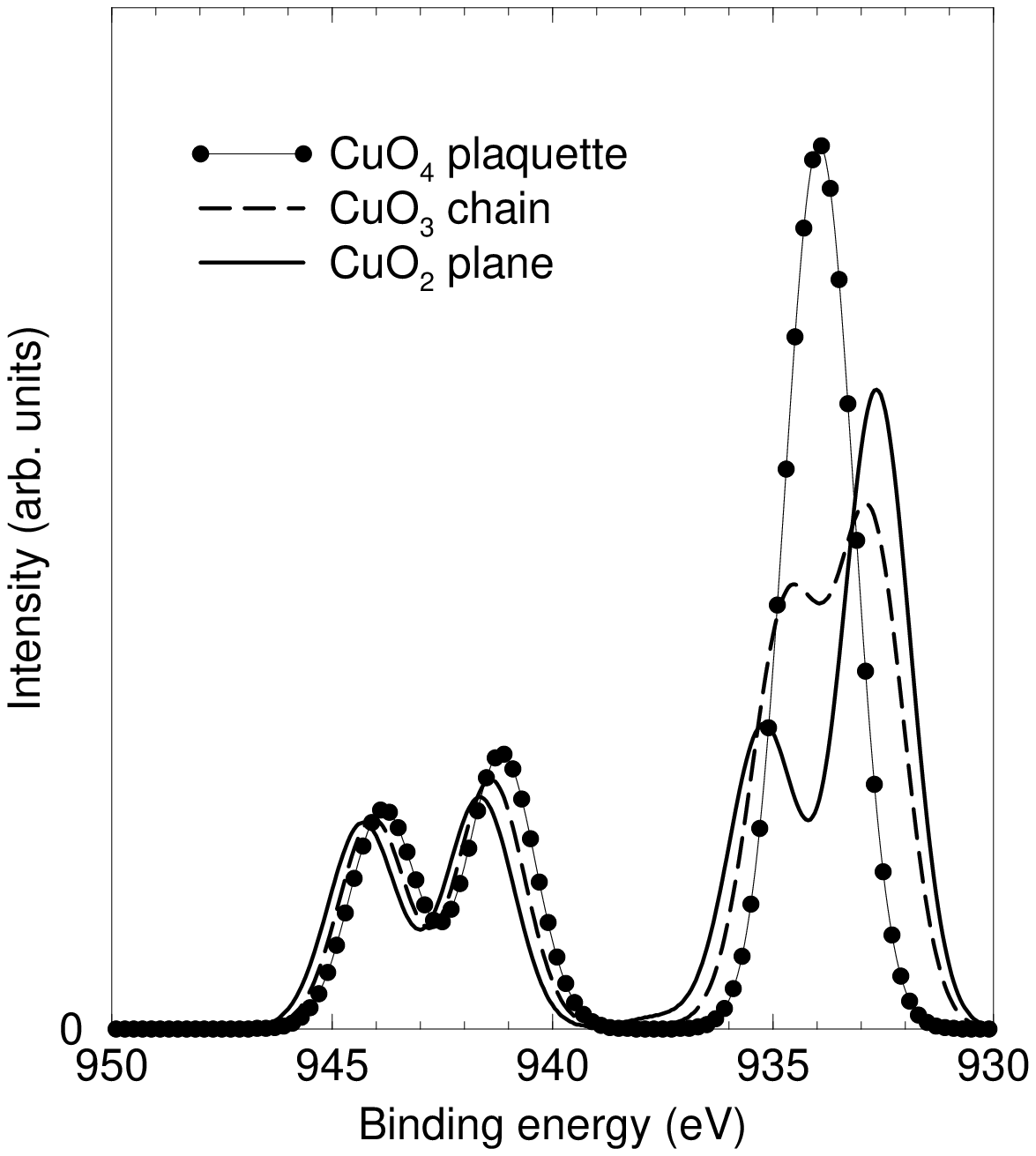}
\begin{figure}
\narrowtext
\caption{Effects of Cu-O network dimensionality. Since the spectra of 
the three different Cu-O networks are shown using the same parameter set 
(\ref{Values}) in all cases, all differences are exclusively due to
dimensionality effects. The line width is $\Gamma =1.8~\mbox{eV}$, 
while $U_{dc}=8.5~\mbox{eV}$, and $I_{dc}=-1.5~\mbox{eV}$. 
The satellite to main line intensity ratio $I_{s}/I_{m}$ is $0.56$ 
for the CuO$_{4}$ plaquette, $0.50$ for the CuO$_{3}$ chain, 
and $0.49$ for the CuO$_{2}$ plane.}
\end{figure}

The influence of Cu-O network dimensionality on the Cu $2p_{3/2}$ spectra 
is illustrated in Fig.~6 where we show results of the projection technique 
for zero-, one- and two-dimensional Cu-O networks. Since the same parameter 
set (\ref{Values}) has been used for all geometries, all changes in the 
spectra are exclusively due to dimensionality effects. 
The most important effect in the spectra is that for the 
one- and two-dimensional system an additional excitation appears at lower 
binding energies. As discussed above the final states of this excitation 
are delocalized and rather similar for both the one- and 
the two-dimensional structures, see states 
(c) in Figs.~3 and 5. Overall, there is only a quantitative change 
from one to two dimensions, in contrast to the qualitative change 
observed between zero and higher dimensions. This conclusion is
in principal agreement with results of exact diagonalization 
calculations.\cite{Okada97jesrp} The peak
around $934~\mbox{eV}$ binding energy which dominates the main line in 
the case of zero dimensions becomes a shoulder structure which decreases 
in intensity and shifts towards higher binding energies as the 
dimensionality increases. Nevertheless, the final state associated with 
this peak preserves its main properties (a large valence-hole density at 
the O sites around the core-hole site) with changing dimensionality, 
see states (a) in Figs.~2, 3, and 5. 
As the dimensionality increases the delocalization in the
Cu-O network increases. Therefore one observes a monotonic decrease in the 
ratio $I_{s}/I_{m}$ for increasing dimensions. Since this trend is not 
observed experimentally, the actual value of $I_{s}/I_{m}$ has to depend 
mainly on material-specific properties. In our calculation these properties
are reflected by the values of the model parameters, cf. Figs.~2 to 5.

%%%%%%%%%%%%%%%%%%%%%%%%%%%% SECTION IV %%%%%%%%%%%%%%%%%%%%%%%%%%%%%%%%%%

\section{Summary}

Summing up, we have calculated the Cu $2p_{3/2}$ core-level spectra of 
zero- one- and two-dimensional Cu-O networks using a model Hamiltonian 
which describes exchange splitting and delocalization within one framework. 
The spectral intensity has been obtained using the Mori-Zwanzig projection 
technique. This method leads to the exact solution for a single CuO$_{4}$ 
plaquette, and we observe excellent convergence in the case of an infinite 
CuO$_{3}$ chain and an infinite CuO$_{2}$ plane. The delocalization 
properties of the final states obtained in the calculation are easy to 
interpret. The results have been compared to experimental spectra by 
using either $\Delta $ or $t_{pd}$ as a fit parameter. While there is 
a good agreement between theory and experiment in the case of 
Bi$_{2}$CuO$_{4}$ and Sr$_{2}$CuO$_{3}$, one excitation seems to be missing 
when compared with the spectrum of Sr$_{2}$CuO$_{2}$Cl$_{2}$. An analysis 
of the influence of dimensionality effects, as compared to effects due to 
material-specific properties, indicates that this missing feature may be 
due to orbitals which are not contained in the model. It is also found 
that dimensionality plays a minor role for the satellite-main line 
intensity ratio $I_{s}/I_{m}$ which is mainly determined by the values of 
the model parameters (especially the value of $\Delta$). 

%%%%%%%%%%%%%%%%%%%%%%%%%%%% ACKNOWLEDGEMENTS %%%%%%%%%%%%%%%%%%%%%%%%%%%%

\acknowledgements
Discussions with J.\ Fink, M.~S.\ Golden, A.\ Goldoni, R.~E.\ Hetzel, 
F.\ Parmigiani, and L.\ Sangaletti
are gratefully acknowledged. This work is supported 
by DFG through the research program of the SFB 463, Dresden.

%%%%%%%%%%%%%%%%%%%%%%%%%%%% REFERENCES %%%%%%%%%%%%%%%%%%%%%%%%%%%%%%%%%%

\end{multicols}

\end{document}